\documentclass[journal]{IEEEtran}
\usepackage{units}
\usepackage{ifpdf}
\usepackage{cite}
\ifCLASSINFOpdf
  \usepackage[pdftex]{graphicx}
\else
  \usepackage[dvips]{graphicx}
\fi
\usepackage[cmex10]{amsmath}
\usepackage{mdwtab}
\usepackage[font=footnotesize,caption=false]{subfig}
\usepackage{url}

\newcommand{\abl}[2]{\frac{{\rm d} #1}{{\rm d} #2}}  

\newcommand{\twoCases}[4]{
  \left\{ 
    \begin{array}{ll} 
      #1 & #2 \\
      #3 & #4 
    \end{array} 
  \right.
}
\begin{document}

\title{An Open-Source Microscopic Traffic Simulator}

\author{Martin~Treiber and~Arne~Kesting\thanks{M. Martin and A. Kesting are with the Department of Transport and Traffic Sciences, Technische Universit\"at Dresden, W\"urzburger Str. 35, 01062 Dresden, Germany. e-mail: (see http://www.mtreiber.de and http://www.akesting.de).}}

\markboth{IEEE Intelligent Transportation Systems Magazine~Vol.~2, No.~3, 6-13, 2010}{Treiber and Kesting: An Open-Source Microscopic Traffic Simulator}

\maketitle

\begin{abstract}
We present the interactive Java-based open-source traffic simulator available at \texttt{www.traffic-simulation.de}. In contrast to most closed-source commercial simulators, the focus is on investigating fundamental issues of traffic dynamics rather than simulating specific road networks. This includes testing theories for the spatiotemporal evolution of traffic jams, comparing and testing different microscopic traffic models, modeling the effects of driving styles and traffic rules on the efficiency and stability of traffic flow, and investigating novel ITS technologies such as adaptive cruise control, inter-vehicle and vehicle-infrastructure communication. 
\end{abstract}


\section{\label{sec:intro}Introduction}
\begin{figure}
\centering \includegraphics[width=0.46\textwidth]{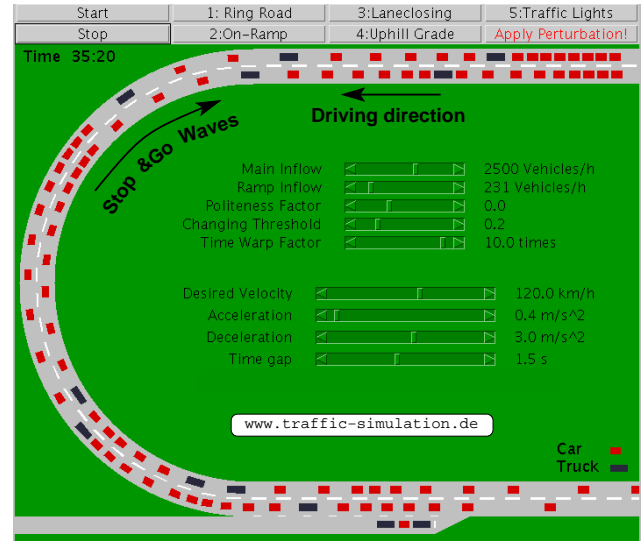}

\caption{\label{fig:simulator}Screenshot of website \texttt{www.traffic-simulation.de}. Shown is the on-ramp scenario with the 
ramp serving as a bottleneck and producing stop-and-go waves propagating against the driving direction. The Java source code of the simulator is publicly available as open source.}
\end{figure}

\IEEEPARstart{I}{}n  the late 1990s, the
simulator presented in this contribution started off as an
educational project. Later on, a version of it was made publicly
available  in form of the Internet applet at the website \texttt{www.traffic-simulation.de}.
Besides interactive simulations of various
predefined traffic situations (see Fig.~\ref{fig:simulator} for a
screenshot), this site discusses the applied
microscopic traffic models \cite{Opus,MOBIL-TRR07} and offers open access to the Java source code.

\begin{figure}
\centering \includegraphics[width=0.46\textwidth]{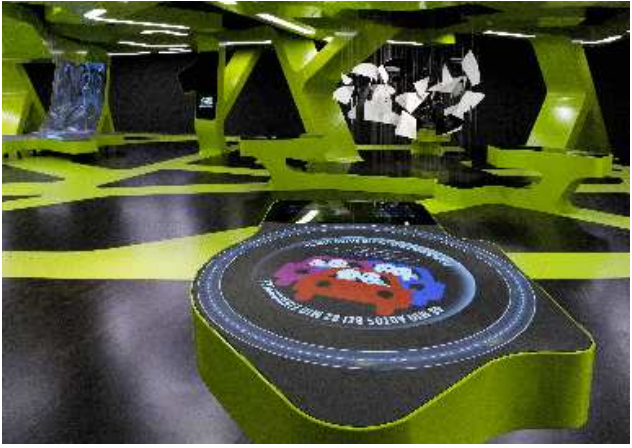}
\caption{\label{fig:levelGreen}Public exhibit demonstrating economic
driving (exhibition ``Level Green'', \textit{Volkswagen Autostadt},
Wolfsburg). Shown are traffic instabilities on a ring-road which can
be influenced interactively by the visitors.}
\end{figure}

In spite (or just because) of its simplicity, the applet became
popular in the course of time. It was displayed, e.g., on the German
computer exhibition CeBIT 2009 and 2010, in the Wall Street Journal (July 1, 2005), and
in several radio and TV broadcasts (see also Fig.~\ref{fig:levelGreen}).

As primary purpose, the simulator demonstrates the
dynamics of traffic flow to the general public. For example, when
simulating the ``ring-road''  and selecting intermediate
densities, one can observe traffic
instabilities eventually leading to stop-and-go waves.
When choosing real open systems (scenarios
2-4), however, one needs
three building blocks to cause a jam: 
(i)~high traffic demand, (ii)~a bottleneck, and (iii)~a perturbation
in the traffic flow~\cite{Treiber-ThreePhasesTRB}. The traffic demand can be controlled by sliders
(``main inflow'' and ``ramp inflow'') while the  bottleneck (onramp, lane
closing or an uphill grade) is selected by simulating the
appropriate scenario. To enhance the breakdown, perturbations
to the traffic flow can 
be added by letting one vehicle brake without reason.\footnote{Since
further perturbations are inherently provoked by lane changes,
traffic can also break down without external action.}
It is instructive to see that the car actually causing the
breakdown (marked with a different color in the simulation) is not
affected by it. The driver is likely not even aware of 
the consequences of his or her maneuver because the breakdown takes
several minutes to develop.\footnote{The default parameter settings correspond
to extremely unstable traffic. For more realistic
values, it would take even longer~\cite{Treiber-Verkehrsdynamik}.} 

The computer program can also be used to simulate the effects of roadside
control. For example, running the lane-closing scenario with the
default settings will result in a traffic breakdown at the
bottleneck. However, imposing a speed limit of \unit[80]{km/h} will
prevent the breakdown. In this case, cars and trucks drive
at nearly the same (desired) speed and lane changes from the lane to be
closed can be performed comparatively easy. 
This reflects the \emph{golden rule} 
of traffic-flow optimization stating that perturbations in the traffic flow
should be minimized~\cite{Arne-ACC-TRC}. Finally, the effects of
different driving styles can be demonstrated: ``Agile'' drivers
(which are characterized by high values of the acceleration parameter)
dampen traffic waves and help dissolving the jam. In contrast,
non-anticipative drivers (braking at the very last moment and
characterized by a high value of the deceleration
parameter) destabilize traffic flow~\cite{kesting-acc-roysoc}.

Extensions of the Internet simulator, mainly targeted for  car and traffic
engineers, are not yet available on the website but
fully operating and documented. They include
\begin{itemize}
\item a simulation testbed for assessing the performance of a wide variety of
models in standard scenarios~\cite{Treiber-Verkehrsdynamik}. This includes
time-continuous models, iterated coupled maps, cellular
automata, and mixtures thereof.
\item simulating varying percentages of vehicles equipped with
driving-assistance systems such as semi-automated driving
(adaptive cruise control, ACC)~\cite{Arne-ACC-TRC,kesting-acc-roysoc}
\item A basis for simulating the performance and efficiency of various
protocols for car-to-car (C2C) and car-to-infrastructure (C2I)
communication as a function of the percentage of equipped
vehicles~\cite{Kesting-IVC-Transactions09,thiemann-IVC-PRE08}.
Furthermore, the simulator allows testing the 
performance of C2C/C2I applications such 
as local online traffic-state recognition~\cite{IVC-Martin-TRR07}.
\item a microscopic physics-based modal model for calculating fuel consumption and
emissions in various dynamic traffic conditions
(cf. Figs.~\ref{fig:vla} and~\ref{fig:levelGreen}),
\item the generation of realistic background traffic flow for scientific
driving simulators (as opposed to computer games).
\end{itemize}

The outline is as follows. After describing the simulator and its
acceleration and lane-changing models in Sec.~\ref{sec:models},
we will describe how to simulate traffic jams in
different scenarios in Sec.~\ref{sec:stylizedFacts}.
Section~\ref{sec:applications} is devoted to applications such as
simulating fuel consumption and driver assistance systems.
Sec.~\ref{sec:outlook} concludes with an outlook.

\section{\label{sec:models}Simulator Description}

The simulator is based on the microscopic modeling approach
describing the movement  of each vehicle or, equivalently, the action of each driver.
\begin{itemize}
\item The \textit{longitudinal dynamics} (accelerating and braking behavior) is
described by a time-continuous 
car-following model (Sec.~\ref{sec:IDM}).
\item The \textit{transversal dynamics} (lane-changing behavior) is described by a discrete-choice
model which is consistent with the longitudinal model (Sec.~\ref{sec:MOBIL}).
\end{itemize}

\subsection{\label{sec:IDM}Longitudinal movement}
Accelerations and decelerations are modeled with the Intelligent
Driver Model (IDM)~\cite{Opus}. The IDM acceleration is
a continuous function of 
the (bumper-to-bumper-) distance $s$ to the leading
vehicle, the speed $v$ of the vehicle considered, and the speed difference
(approaching rate) $\Delta v=v - v_\text{l}$ to the leading vehicle
$l$. It is given by
\begin{eqnarray}
\label{IDM}
\frac{\text{d}v}{\text{d} t} &=&  a
         \left[ 1 -\left( \frac{v}{v_0} \right)^\delta -\left(
         \frac{s^*(v,\Delta v)} {s}
         \right)^2\, \right], \\
\label{sstar}
s^*(v, \Delta v)  &=& s_0  + v T + \frac{v \Delta v }  {2\sqrt{a b}}.
\end{eqnarray}
The IDM acceleration combines the free-road acceleration strategy
$a_\text{free} (v)= a[1-(v/v_0)^\delta]$ with a deceleration strategy
$a_\text{brake}(s, v, \Delta v) = -a(s^*/s)^2$ 
that becomes relevant when the gap to the leading vehicle is not
significantly larger than the effective ``desired'' (safe) gap
$s^*(v, \Delta v)$.
The free acceleration is characterized by the
\textit{desired speed} $v_0$, the
\textit{maximum acceleration} $a$, and the exponent $\delta$
characterizing how the acceleration decreases  with the speed
($\delta=1$ corresponds to a linear decrease while $\delta\to \infty$
corresponds to a constant acceleration). The effective minimum gap $s^*$
is composed of the \textit{minimum distance} $s_0$ (which is relevant
for low velocities only), the speed-dependent distance $vT$ which corresponds
to following the leading vehicle with a constant \emph{desired time
gap} $T$, and  a dynamic contribution which is only active in non-stationary traffic
corresponding to situations in which $\Delta v\ne 0$. 
This latter contribution implements an
\emph{intelligent} driving behavior that, in normal situations,
 limits braking decelerations to the
\emph{comfortable deceleration} $b$. In critical situations,
however, the IDM deceleration  becomes significantly higher,
making the IDM \emph{collision-free}~\cite{Opus}. Notice that the
IDM acceleration function~\eqref{IDM} is continuous and the different driving
situations are  connected by implicit smooth transitions, not by
explicit conditions.

The IDM parameters
$v_0$, $a$, $b$, $T$, and $s_0$ (the first four can be changed
interactively, see Fig.~\ref{fig:simulator})
have a reasonable interpretation, are known to be relevant,
are empirically measurable, and have realistic
values~\cite{Kesting-Calibration-TRR08}. 
Furthermore, the model is complete in the sense that it describes all
traffic situations (such as free and congested traffic including all
transitions) on freeways as well as in cities. For two situations, it has two minor
imperfections:
\begin{enumerate}
\item IDM drivers overreact in response to very small gaps that may
be caused by ``cut-in'' lane-changing maneuvers of other drivers. This
is relevant for ACC applications and for realistic lane-changing
behavior. A modification resolving this issue is proposed 
in Ref.~\cite{kesting-acc-roysoc}.
\item In car-following situations near the desired speed, the
equilibrium time gap becomes significantly larger than the time-gap parameter
$T$. This is undesirable in ACC implementations where the time gap can
be set by the driver (``you should get what you want''). 
Moreover, an unrealistic spreading is observed
in some platooning situations. A solution consists in replacing Eq.~\eqref{IDM} with
\begin{equation}
\label{IIDM}
\abl{v}{t}=\twoCases{a\left[1-\left(\frac{s^*}{s}\right)^2\right]}{\ \frac{s^*}{s}\ge 1,}
{a_\text{free}  \left[1-\left(\frac{s^*}{s}\right)
 ^{\frac{2a}{a_\text{free}}}\right]}{\ \text{otherwise.}}
\end{equation}
\end{enumerate}
Both modifications neither change the parameter set nor the dynamics
in other situations nor the continuity of the acceleration function. Nevertheless,
for the sake of simplicity, we will retain the original
IDM acceleration~\eqref{IDM} with Eq.~\eqref{sstar}.\footnote{Moreover, some
drivers do increase their gap when 
following other vehicles near the desired speed. For these drivers, the
original IDM is even more realistic than the
modification~\eqref{IIDM}.} In future versions, it will be possible to
chose between several models and variants.

\subsection{\label{sec:MOBIL}Lane-Changing Model MOBIL}
Like the acceleration model, the lane-changing model is selected
according to the Einstein postulate which can be paraphrased by
\textit{make it as simple as possible -- but not
simpler}. Consequently, all the complicated strategic and tactical
preparation stages are discarded focussing exclusively on the operational
(ad-hoc) lane-changing decision. Furthermore, complicated traffic
rules (such as forbidding overtaking on the right lane)
are ignored. However, speed differences play a
crucial role and must be retained as explanatory variable 
(gap-acceptance models where lane-changing
decisions are just based on speed-dependent gaps are not
sufficient). Moreover, the lane-changing decision must be compatible 
with the acceleration model. In particular, the combined models should
be accident-free if this is the case for the longitudinal model alone. In
commercial simulators this is achieved by nesting the two submodels
to a  complex compound acceleration and lane-changing model.

Remarkably, this complication can be avoided by simple
acceleration-based lane-changing rules~\cite{MOBIL-TRR07}. 
 For a vehicle $c$ considering a lane change,
the subsequent vehicles on the target (new) and present (old) lanes are represented
by $n$ and $o$, respectively. The acceleration $a_c$ denotes the
(IDM) acceleration of vehicle $c$ on the actual lane, while $\tilde{a}_c$
refers to the prospective situation after a completed change, i.e., to the new
acceleration of vehicle $c$ on the target lane. Likewise,
$\tilde{a}_o$ and $\tilde{a}_n$ denote the acceleration of the old and
new followers after completion of the lane change.

The set of lane-changing rules can be comprised by two criteria.
The \emph{safety criterion} checks the possibility of executing a lane change
by allowing only changes where the  subsequent vehicle on the target
lane is not forced to brake at a deceleration exceeding a 
given safe limit $b_\text{safe}$, 
\begin{equation}\label{safety} 
\tilde{a}_n \ge -b_\text{safe}.
\end{equation} 
Although formulated as a simple inequality,  this condition contains
all the information provided by the acceleration model. In particular,
 the admissible
minimum time gap may be significantly larger than $T$ (fast target
vehicle, $v_n>v_c$), or significantly smaller (slower target vehicle).

The \emph{incentive criterion} determines if a lane change is
desirable. Like the safety criterion, ``desirability'' is based on the
accelerations of the longitudinal model before and after the
prospective lane change.
Furthermore,  we generalize the incentive criterion to include the
immediately affected neighbors as well. The \textit{politeness factor}
$p$ determines to which degree these vehicles influence the
lane-changing decision (cf. Fig.~\ref{fig:simulator}). 
For symmetric overtaking rules, we neglect
differences between the lanes and propose the following incentive
condition for a lane-changing decision of the driver of vehicle $c$:

\begin{equation}
\label{eq:MOBIL}
\underbrace{\tilde{a}_c-a_c}_{\text{driver}} 
+ p \big( \underbrace{\tilde{a}_n-a_n}_{\text{new follower}}
+ \underbrace{\tilde{a}_o-a_o}_{\text{old follower}} \big) > \Delta a_\text{th}.
\end{equation}

The first two terms denote the advantage (\textit{utility}) of a possible lane
change for the driver of the considered vehicle $c$. 
The third term with the {\it politeness factor} $p$ 
denotes the total advantage (acceleration gain or loss, if
negative) of the two immediately affected neighbors, weighted with
$p$.\footnote{Most other lane-changing models correspond to setting
$p=0$.}
Finally, the switching threshold $\Delta a_\text{th}$ on the
right-hand side of Eq.~\eqref{eq:MOBIL} models a certain
inertia and prevents lane changes if the overall advantage is only
marginal compared to a ``keep lane'' directive. Notice that
the incentive criterion~\eqref{eq:MOBIL} acts simultaneously as a
\emph{safety} criterion for 
the lane-changing vehicle $c$ (in contrast to the safety
criterion~\eqref{safety} ensuring the safety of vehicle $n$). In order
to avoid erroneous decisions for the case of two vehicles driving in
parallel, the acceleration model must return a prohibitively negative
acceleration for this case (corresponding to negative gaps). Finally, asymmetric
traffic regulations such as keep-right directives can be implemented
by an additional bias term $\pm a_\text{bias}$ (positive for changes
to the left) on the right-hand side
of Eq.~\eqref{eq:MOBIL}~\cite{MOBIL-TRR07}.

\section{\label{sec:stylizedFacts}Simulating Empirical Phenomena}
Traffic congestions cannot evolve arbitrarily
in space and time. Instead, observations on freeways 
all over the world indicate that the dynamics of traffic jams obeys
certain empirical rules, also known as \emph{stylized
facts}~\cite{Treiber-ThreePhasesTRB}. Taking them into account can
greatly improve the  traffic
state estimation from sparse data~\cite{ASM-CACAIE}. Some of the facts
that can be seen in Fig.~\ref{fig:stylizedFacts} (top) are the
following (for a complete list, see Ref.~\cite{Treiber-ThreePhasesTRB}):

\begin{figure}
\centering \includegraphics[width=0.46\textwidth]{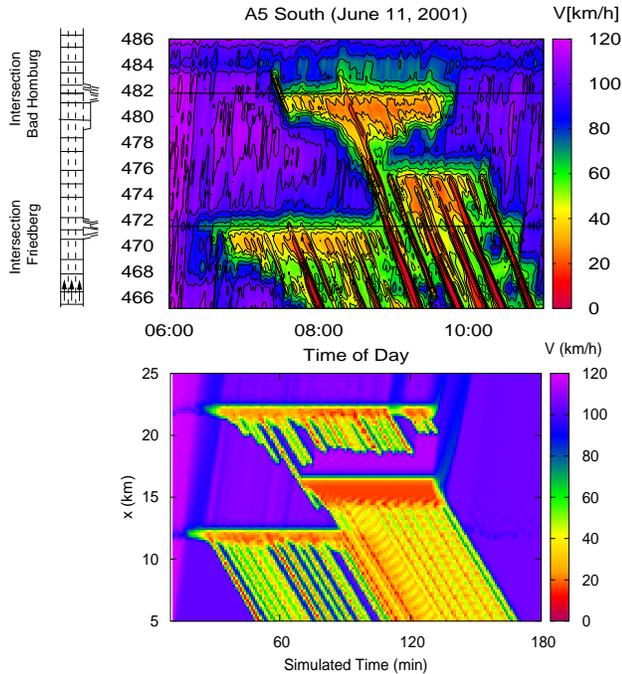}
 \caption{\label{fig:stylizedFacts}Spatiotemporal evolution of traffic
jams caused by two intersection and a temporary bottleneck. Shown
(top) is the speed estimated by a dedicated method~\cite{ASM-CACAIE} from aggregated
cross-sectional data at positions depicted in the sketch on the
left-hand side. Also shown is a simulation with the Intelligent Driver Model (bottom). }
\end{figure}

\begin{itemize}
\item
Nearly all breakdowns are caused by bottlenecks in combination with a perturbation in the
traffic flow. In Fig.~\ref{fig:stylizedFacts}, the bottlenecks are
constituted by two intersections, and by a temporary lane closing (accident)
near road kilometer 476. Other types include junctions (scenario 2 in
the Internet simulator), road works (scenario 3), or gradients (scenario 4).
\item
The downstream front (where cars leave the congestion) is either
stationary at the bottleneck, or moves against the driving direction with a characteristic
velocity $c_{\rm cong}$ of about~\unit[-15]{km/h}. Both cases can occur
in a single jam (see Fig.~\ref{fig:stylizedFacts} at road kilometer~\unit[476]{km} at time \mbox{10:10}).  
\item
Jams can be localized or extended. Most extended jams show internal
structures propagating all at the same velocity $c_{\rm cong}$. Furthermore,
they grow in
amplitude eventually resulting in
stop-and-go waves or even isolated moving jams (perceived as ``phantom
jams'' by drivers). 
\end{itemize}

\noindent
In the simulated speed field shown in 
Fig.~\ref{fig:stylizedFacts} (bottom), the simulator is set up with
upstream and downstream boundary conditions from the corresponding
detectors. The two permanent bottlenecks are simulated by simple
on-ramps while the temporary lane closing is modeled by a
``flow-conserving bottleneck'': The model parameters $v_0$ and $T$ are
changed locally and temporally such that the effective road capacity
is reduced in the corresponding spatiotemporal region. The
simulation reproduces the \emph{complete set} of stylized facts with just the
boundary conditions and the arrangement of bottlenecks as input. However, the 
agreement with the data is not quantitative. Particularly, the
wavelengths of the simulated stop-and-go waves are too
short. This could be resolved by more elaborate models~\cite{HDM}. 


\section{\label{sec:applications}Applications}

Intelligent traffic systems (ITS) such as adaptive cruise control
(ACC) or car-to-car and car-to-infrastructure communication (C2C/C2I)
are becoming increasingly widespread and begin to influence traffic
operations. This opens up new vehicle-based possibilities to optimize the efficiency
of traffic flow. The objective function to be maximized typically contains
savings in traveling time or fuel consumption/CO$_2$ emissions. 
In the following subsections, we show how our software can be applied
to calculate fuel consumptions, and to simulate the effects of 
novel ITS systems. Since these systems do not yet exist, simulation is
the only means to assess their properties.

\subsection{\label{sec:fuel}Fuel Consumption}

When discussing how much a specific traffic situation (particularly,
jams) and the driving 
style (including automated driving) influence fuel consumption and
emissions,  one needs a detailed microscopic (modal) consumption model. 
Our simulator implements the physics-based model described 
in Ref.~\cite{Treiber-Fuel-TRB08}. This model contains 
intuitive and measurable parameters such as
mass, friction and wind-drag coefficients of the respective
vehicle. Moreover, 
 its exogenous variables, namely speed $v$
and acceleration $\dot{v}$, are just the output variables of the
actual dynamic simulation. Specifically, the consumption rate
$\dot{c}$ (liters per time unit) of a vehicle is given by
\begin{equation}
\label{dotc}
\dot{c}(v,\dot{v})=\frac{P}{\gamma(P,f) w_{\text{cal}}}
\end{equation}
where the required total power $P$ depends, via a physical formula, on 
speed, acceleration, road gradient, and some vehicle
properties~\cite{Treiber-Fuel-TRB08}. Furthermore, the fraction $\gamma$ of the fuel calorific
energy (given per volume by $w_{\text{cal}}$)  which is converted into mechanic
energy depends,
via the engine characteristics, on $P$ and the rotation rate $f$. 
The rotation rate, in turn, depends on the speed and the chosen gear.

\begin{figure}
\centering \includegraphics[width=0.46\textwidth]{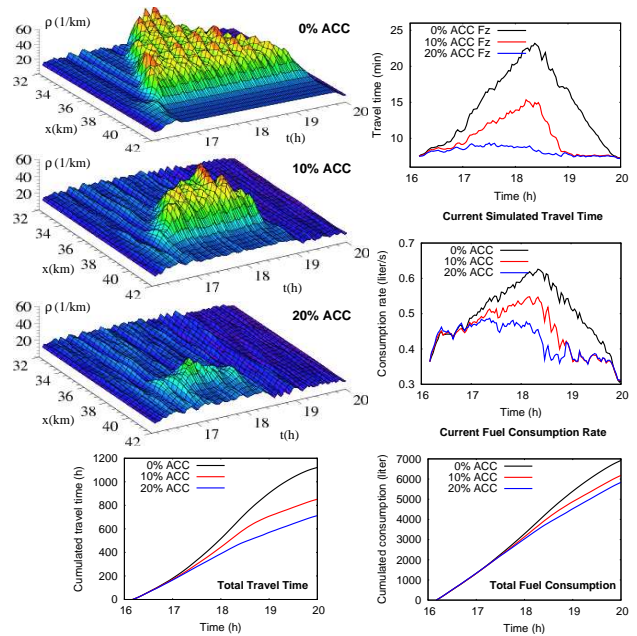}
  \caption{\label{fig:vla}Influence of adaptive driving behavior on
traffic flow. Shown are simulation results for mixed traffic with \unit[0]{\%}
(top), \unit[10]{\%} (middle), 
and \unit[20]{\%} (bottom) of cars implementing the adaptive driving strategy (see the
main text for details).}
\end{figure}

Figure~\ref{fig:vla} shows an application of the fuel-consumption
module in three situations with variable amounts of
jammed traffic. Assumed are a level road,  
a certain type of middle-class vehicle, and a fuel-efficient gear
choice for a given speed and acceleration.
We obtain the remarkable result that the relative savings in travel
time by jam avoidance are at least three times higher than that in
consumption/emissions. 

\subsection{\label{sec:vla}Adaptive Driving Behavior}

As discussed in Sec.~\ref{sec:stylizedFacts}, one
 ``building block'' for traffic breakdowns are bottlenecks, i.e., 
road sections with locally decreased capacity. However, the capacity,
i.e., the maximum possible throughput (vehicles per hour), does not
only depend on the infrastructure (number of lanes etc.) but also on
the driving style. Consequently, by local changes of the driving
style, it is possible to \textit{dynamically fill the capacity gaps}
of the bottlenecks, at least partially. This is the rationale behind
the proposed adaptive driving
behavior~\cite{Arne-ACC-TRC,kesting-acc-roysoc} which can be
implemented 
either by automated driving (``traffic-adaptive ACC''\footnote{This
corresponds to a ``doubly adaptive'' cruise control adapting to the
immediate predecessor as well as to the local traffic situation.}),
or by 
hints about effective driving which could be taught in driving lessons.

\begin{figure*}
\centering \includegraphics[width=0.65\textwidth]{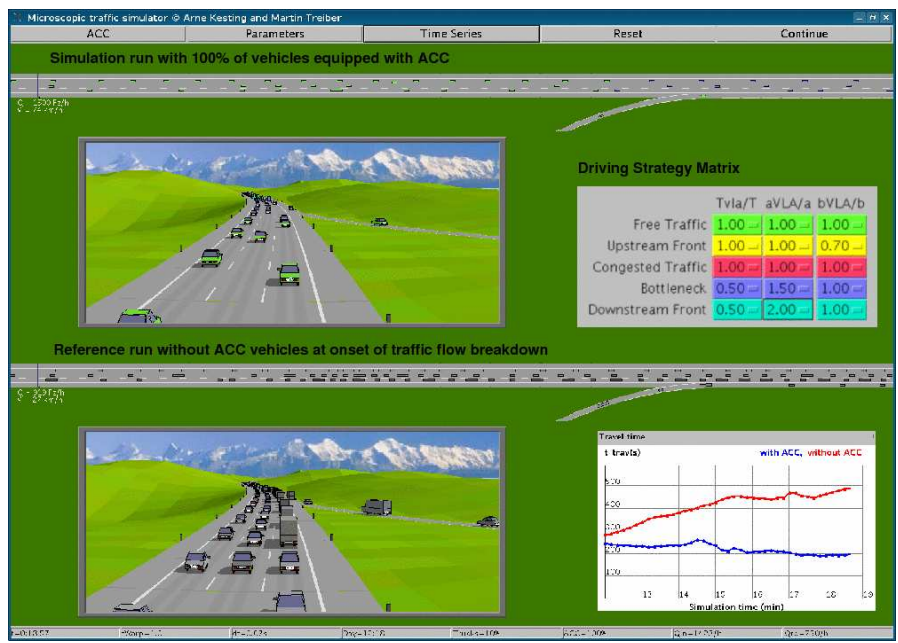}
  \caption{\label{fig:vla3d}Screenshot of the simulator while
simulating the effect of adaptive driving behavior~\cite{Arne-ACC-TRC}. The driving style
as a function of the  traffic situation is controlled by the ``Driving Strategy Matrix''.
}
\end{figure*}

Adaptive driving is implemented in our simulator by the ``strategy
matrix'' coupling a set of driving situations to a set of actions
(cf. Fig.~\ref{fig:vla3d}). The
five driving situations (free and congested traffic, jam entering and
leaving, driving through a bottleneck) relate
to traffic situations as well as to the infrastructure. The actions,
i.e., changes of the driving style, are modeled by changes of the IDM
parameters $a$, $b$, and $T$. Here it comes in handy that these
parameters are intuitive and correspond to nearly orthogonal aspects
of the driving style: 
\begin{itemize}
\item The \emph{agility} of the drivers is positively correlated to the acceleration
parameter $a$,
\item the \emph{degree of anticipation} is negatively correlated to
the comfortable braking deceleration as discussed in
Sec.~\ref{sec:IDM},
\item and the \emph{following distance} (aggressiveness if too
close) is determined by the desired time headway $T$.
\end{itemize}
Notice that the set of actions does not include changes of the desired
velocity $v_0$ since this would reflect the effect of roadside
control (speed limits) rather than in-vehicle traffic optimization.

The complete adaptation strategy is encoded by the fifteen matrix
elements denoting the relative change of the parameter values with
respect to the default driving style.\footnote{Strictly speaking, only
twelve
elements are varied since the values for ``free traffic'' can be identified with
the default set.} The matrix elements are given in relative
terms. Thus, changes of the driving style are orthogonal to
inter-driver variations representing the heterogeneity of drivers and
vehicles.\footnote{For example, an increase in agility applied to a
sluggish driver means that he or she just becomes a little bit less sluggish.}
The strategy represented by the matrix of Fig.~\ref{fig:vla3d} correspond
to comparatively modest
actions (only five elements are different from unity) which,
nevertheless, lead to a 
significant increase of efficiency. This strategy can be expressed by
following rules: 
\begin{itemize}
\item Braking early when  approaching a jam
(row 2),
\item increasing the agility and reducing gaps (of course, without violating
safety limits)  at bottlenecks (row 4),
\item closely following the leader and accelerating fast
when leaving a jam (row 5).
\end{itemize}
Figure~\ref{fig:vla} shows three simulations with identical traffic
volume where \unit[0]{\%}, \unit[10]{\%}, and \unit[20]{\%} of
all vehicles are equipped with traffic-adaptive ACC (or driven by
traffic-aware drivers). 
With respect to capacity, this strategy leads to an effective sensitivity
$\epsilon=0.3$, i.e., \unit[1]{\%} more equipped vehicles will increase the capacity by
about \unit[0.3]{\%}. However, with respect to maximum travel times (TT) and
consumption rates (C), the
nonlinearities associated with jams constitute a significant
leverage effect resulting (in this specific situation) in  $\epsilon_\text{TT} \approx 4$ and
$\epsilon_C \approx 1$.

\subsection{\label{sec:ivc}Car-to-Car and Car-to-Infrastructure Communication}
\begin{figure*}
\centering \includegraphics[width=0.8\textwidth]{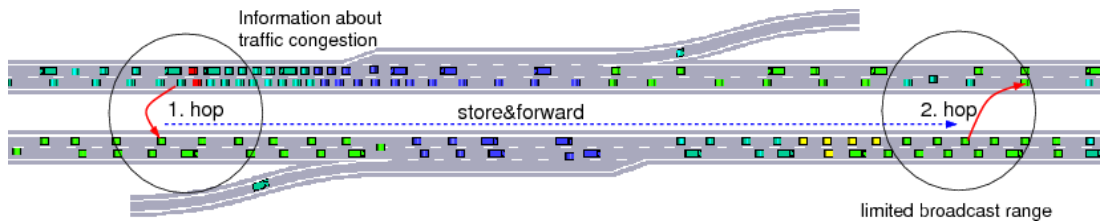}
 \caption{\label{fig:ivc_hopping}Illustration of the store-and-forward
 strategy using the opposite driving direction for propagating
 messages via short-range inter-vehicle communication in upstream
 direction. First, a message is generated on the occasion of a local
 change in speed. The broadcasted message will be picked up by an
 equipped vehicle in the opposite driving direction (first hop). After
 a certain traveling distance, the vehicle starts broadcasting the
 message which can be received by vehicles in the original driving
 direction (second hop).}
\end{figure*}

\begin{figure}
\centering \includegraphics[width=0.4\textwidth]{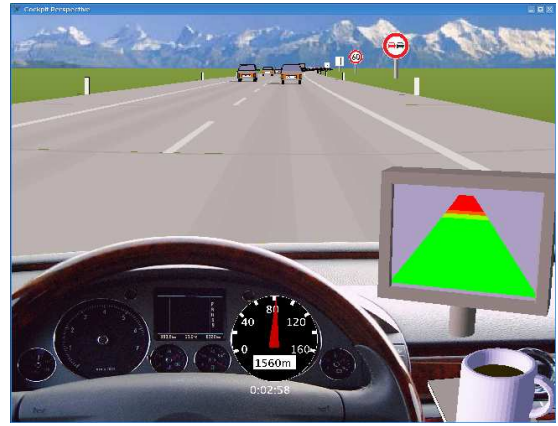}
 \caption{\label{fig:BAS}Simulation of an experimental C2I-based
assistance system providing an online estimation of the local traffic environment near road-work
bottlenecks. In the screenshot, congested traffic is detected ahead
(depicted in red in the stylized ``navigation system'')  that is not yet visible
from the driver's perspective. Also shown (bottom right) is the
``coffeemeter'' visualizing the driving comfort.}
\end{figure}

%
Traffic-adaptive driving strategies such as the one presented in the
previous subsection are only effective if the local traffic situation
is known. For example, to identify the traffic situations
corresponding to the rows 2 and 5 of the strategy matrix, 
the precise locations of the upstream and downstream
fronts of congestions must be known.
Since, from the driver's perspective (Fig.~\ref{fig:BAS}),
one cannot distinguish between jam fronts and spurious oscillations
(Fig.~\ref{fig:stylizedFacts}), the information must be provided by
another assistance system. 
Figure~\ref{fig:ivc_hopping} sketches a possible solution using
store-and-forward car-to-car communication between equipped
vehicles. In order to be effective for low percentages of equipped
vehicles, cars driving in the opposite direction are used as
information carriers. Theoretical
calculations show that this concept is operative for equipment rates as low as
\unit[2]{\%}~\cite{Kesting-IVC-Transactions09}. Integrated simulations of the
communication and traffic dynamics using our simulator essentially confirm the
analytical
results~\cite{Kesting-IVC-Transactions09,thiemann-IVC-PRE08,IVC-Martin-TRR07}.

Another variant of local ad-hoc communication is C2I
communication which is particularly effective at known
bottlenecks such as road works. In this setup, the information
transport  is not performed by
``courier vehicles'' but by a set of at least two connected
``road-side units'' (RSU), one at the bottleneck, and one several kilometers
upstream.  Each equipped vehicle records its trajectory and delivers
it to the downstream RSU by local communication (WLAN). After
analyzing it there, relevant information (such as travel times or
positions of jam
fronts) is transmitted to the upstream RSU which, in turn, delivers it to passing
equipped vehicles by local communication. Finally, the
information can be graphically presented by the in-vehicle
navigation system (Fig.~\ref{fig:BAS}).

\section{\label{sec:outlook}Future Projects}

Because of its open architecture and full control, the simulator can
be extended in the following aspects.
On the website {\tt www.verkehrsdynamik.de} we will offer a publicly
available platform to test the quality of existing and user-defined
car-following models  in
urban and freeway scenarios~\cite{Treiber-Verkehrsdynamik}.
Furthermore, quality criteria such as fuel consumption~\cite{Treiber-Fuel-TRB08} or total
travel times  from  both the individual and system perspective  will be
made available in a future version.
A unique
feature of the 3D version of our simulator is an intuitive
visualization of the ``driving comfort''. We represent this quantity which is rather elusive to
define in mathematical terms, by a cup of virtual coffee which spills
if the driving becomes uncomfortable~\cite{youtubeWeather}. The dynamics of the coffee
surface is defined by a twodimensional driven damped pendulum~\cite{Kesting-Agents08} taking
into account longitudinal and transversal  acceleration and jerk (the time derivative of the
acceleration) which all contribute to the subjective impression of
discomfort. 


\end{document}